\documentclass{moriond}

\usepackage{xspace,amsmath,amssymb}

\def\araa{{Ann.~Rev.~Astron.~Astrophys.}}
\def\apj{{Astrophys.~J.}}\def\apjl{{Astrophys.~J.~Lett.}}
\def\apjs{{Astrophys.~J.~Supp.}}

\def\aapr{{Astron. Astrophys. Rev.}}

\def\mnras{{Mon.~Not.~R.~Astron.~Soc.}}

\def\prd{{Phys.~Rev.~D}}

\def\prx{{Phys.~Rev.~X}}
\def\prl{{Phys.~Rev.~Lett.}}

\def\physrep{{Phys.~Rep.}}

\def\lrr{{Living Rev. Relativ.}}

\def\natastro{{Nat. Astron.}}

\usepackage[dvipsnames, usenames]{xcolor}
\definecolor{linkcolor}{rgb}{0.0,0.3,0.5}
\definecolor{dodgerblue}{HTML}{1E90FF}
\usepackage[unicode, colorlinks=true, linkcolor=linkcolor, citecolor=linkcolor, filecolor=linkcolor,urlcolor=linkcolor, pdfusetitle]{hyperref}
\usepackage{orcidlink}

\def\be{\begin{equation}}
\def\ee{\end{equation}}
\def\bea{\begin{eqnarray}}
\def\eea{\end{eqnarray}}

\def\qluster{\textsc{qluster}\xspace}

\newcommand{\ssim}{\mathchar"5218\relax\,}

\newcommand{\Photo}{}

\begin{document}

\vspace*{4cm}

\title{\large {\sc qluster}: quick clusters of merging binary black holes}

\author{Davide Gerosa$\,$\orcidlink{0000-0002-0933-3579}$\,^{1,2,3}$, Matthew Mould$\,$\orcidlink{0000-0001-5460-2910}$\,^3$}

\address{ \vspace{0.2cm}
 \footnotesize
$^1$ Dipartimento di Fisica ``G. Occhialini'', Universit\'a degli Studi di\\Milano-Bicocca, Piazza della Scienza 3, 20126 Milano, Italy \\
$^2$ INFN, Sezione di Milano-Bicocca, Piazza della Scienza 3, 20126 Milano, Italy \\
$^3$ School of Physics and Astronomy \& Institute for Gravitational Wave Astronomy,\\University of Birmingham, Birmingham, B15 2TT, United Kingdom}

\maketitle\abstracts{
This short document
 illustrates \qluster: a toy model for populations of binary black holes in dense astrophysical environments. \qluster is a simple
tool to investigate the occurrence and properties of hierarchical black-hole mergers detectable
by gravitational-wave interferometers. \qluster is not meant to rival the complexity of state-of-the-art population synthesis and N-body codes but rather provide a fast, approximate, and easy-to-interpret framework to investigate some of the key ingredients of the problem. These include the binary pairing probability, the escape speed of the host environment, and the merger generation. We also introduce the ``hierarchical-merger efficiency" --- an estimator that quantifies the relevance of hierarchical black-hole  mergers in a given astrophysical environment.
}

\section {(Hierarchical) black-hole mergers}
\label{intro}

Understanding the formation and evolution of stellar-mass binary black holes (BHs) is one of the most outstanding problems in modern astrophysics. Current LIGO/Virgo observations indicate that these objects pair and merge with a rate~\cite{2021arXiv211103634T} of $\ssim 30~{\rm Gpc}^{-3} {\rm yr}^{-1}$. Crucially, BH mergers cannot be explained by gravitational-wave (GW) emission alone. The quadrupole formula implies that objects of  $\ssim 10 M_\odot$ on quasi-circular orbits merge in less than a Hubble time only from separations $\lesssim 10 R_\odot$, which is a few orders of magnitude smaller than the typical orbital separations of binary stars. Explaining the LIGO/Virgo events requires additional additional processes of astrophysical nature able to dissipate enough energy and angular momentum at scales $\gg 10 R_{\odot}$ before GWs become important. %

Mainstream BH-binary formation scenarios exploit either interactions between the two stars of an isolated binary (``field'') or  few-body exchanges in dense environments (``clusters").~\cite{2021hgwa.bookE..16M,2022PhR...955....1M}
For isolated formation, some of the key mechanisms that are commonly invoked  include stable~\cite{2017MNRAS.471.4256V} 
and unstable~\cite{2014LRR....17....3P}
mass transfer as well as chemical mixing induced by tides and rotation.~\cite{2016MNRAS.458.2634M} 
Dynamical formation~\cite{2013LRR....16....4B}
divides
into several plausible astrophysical hosts including young star clusters,~\cite{2010ARA&A..48..431P} 
globular clusters,~\cite{2019A&ARv..27....8G} 
nuclear star clusters,~\cite{2020A&ARv..28....4N}
and accretion disks in active galactic nuclei.~\cite{2008ARA&A..46..475H}
While there is still much debate around the issue, it appears that all formation pathways struggle to reproduce the observed dataset in some regions of the parameter space, pointing to a scenario where
multiple
channels
provide
a non-negligible contribution to the merger rate.~\cite{2021ApJ...910..152Z}

The possible occurrence of hierarchical BH mergers (i.e., BHs that merge multiple times on astrophysically relevant timescales) has %
 emerged as a potential
candidate
to shed some light on the formation-channel debate.~\cite{2021NatAs...5..749G} Indeed, BHs that are formed from previous BHs (and not stars) have distinctive properties:
their masses are $\ssim 95\%$ of the sum of the mass of their progenitor and are thus heavier than each of them, and their dimensionless spins follow a peculiar distribution that peaks at $\ssim 0.7$. %
Hierarchical stellar-mass BH mergers are now a leading explanation for the heavier BHs detected by LIGO/Virgo, notably those in GW190521,~\cite{2020ApJ...900L..13A} whose masses are too large to be explained by conventional stellar evolution. %
Because of such distinctive signatures, there have been numerous attempts to infer if (some of) the observed BHs are hierarchical.~\cite{2021NatAs...5..749G} The fraction of hierarchical mergers in the GW catalog provides a lower limit to the merger rate from dynamical channels, as assembling GW sources via repeated mergers inevitably requires interactions between BHs. 

In this conference contribution we present \qluster (``quick cluster''),~\cite{matthew_mould_2023_7807210} a toy model for the formation of hierarchical BH mergers in dense stellar environments. Ours is a highly simplified approach where we remove most of the astrophysical complexity/richness and only deal with the relativistic aspects of the problem. This is not meant to challenge the predictions of state-of-the-art BH-formation models in astrophysical clusters including e.g. \textsc{cmc},~\cite{2022ApJS..258...22R} \textsc{mocca},~\cite{2013MNRAS.431.2184G}, \textsc{nbody6++gpu},~\cite{2016MNRAS.458.1450W} \textsc{rapster}~\cite{2022arXiv221010055K}, and \textsc{fastcluster}.~\cite{2021MNRAS.505..339M} Rather, we hope our simplification might provide hints on some of the key ingredients of the problem, which can then be properly validated with complete simulations using, e.g., one of the codes above.

\qluster has been used in a few published papers already. The model was first developed by Gerosa and Berti~\cite{2019PhRvD.100d1301G} (2019) to highlight the importance of the escape speed $v_{\rm vesc}$ of the environment where hierarchical mergers take place. Emission of linear momentum via GWs imparts recoils of $\mathcal{O}(100\,{\rm km/s})$ to the remnant of BH mergers. If the escape speed of the cluster is lower than this value, second-generation BHs are unlikely to be retained in the environment. This implies that systems such as globular clusters ($v_{\rm esc}\lesssim 50$ km/s) are intrinsically inefficient at producing hierarchical mergers. %
Gerosa, Giacobbo, and Vecchio (2021)~\cite{2021ApJ...915...56G} used \qluster to show that BHs with masses $\gtrsim 50 M_\odot$ but spins $\lesssim 0.2$ {cannot} be explained by hierarchical mergers and require a different astrophysical setup. \qluster was then exploited by Mould, Gerosa, and Taylor (2022)~\cite{2022PhRvD.106j3013M} as a testbed to develop a machine-learning pipeline that interpolates across astrophysical populations and performs the related statistical inference. \qluster predictions were also used to interpret the event GW190412~\cite{2020PhRvL.125j1103G} and similar implementations were later developed by other groups as well.~\cite{2022arXiv220905766M,2022ApJ...935L..20Z}  

\qluster is now made publicly available at \href{https://github.com/mdmould/qluster}{github.com/mdmould/qluster}.~\cite{matthew_mould_2023_7807210} This document, published in the proceedings of the 57$^{\rm th}$ Rencontres de Moriond -- Gravitation, provides an introduction to the code and a brief exploration of the hierarchical-merger efficiency.

\begin{figure}\centering
\includegraphics[width=0.75\textwidth]{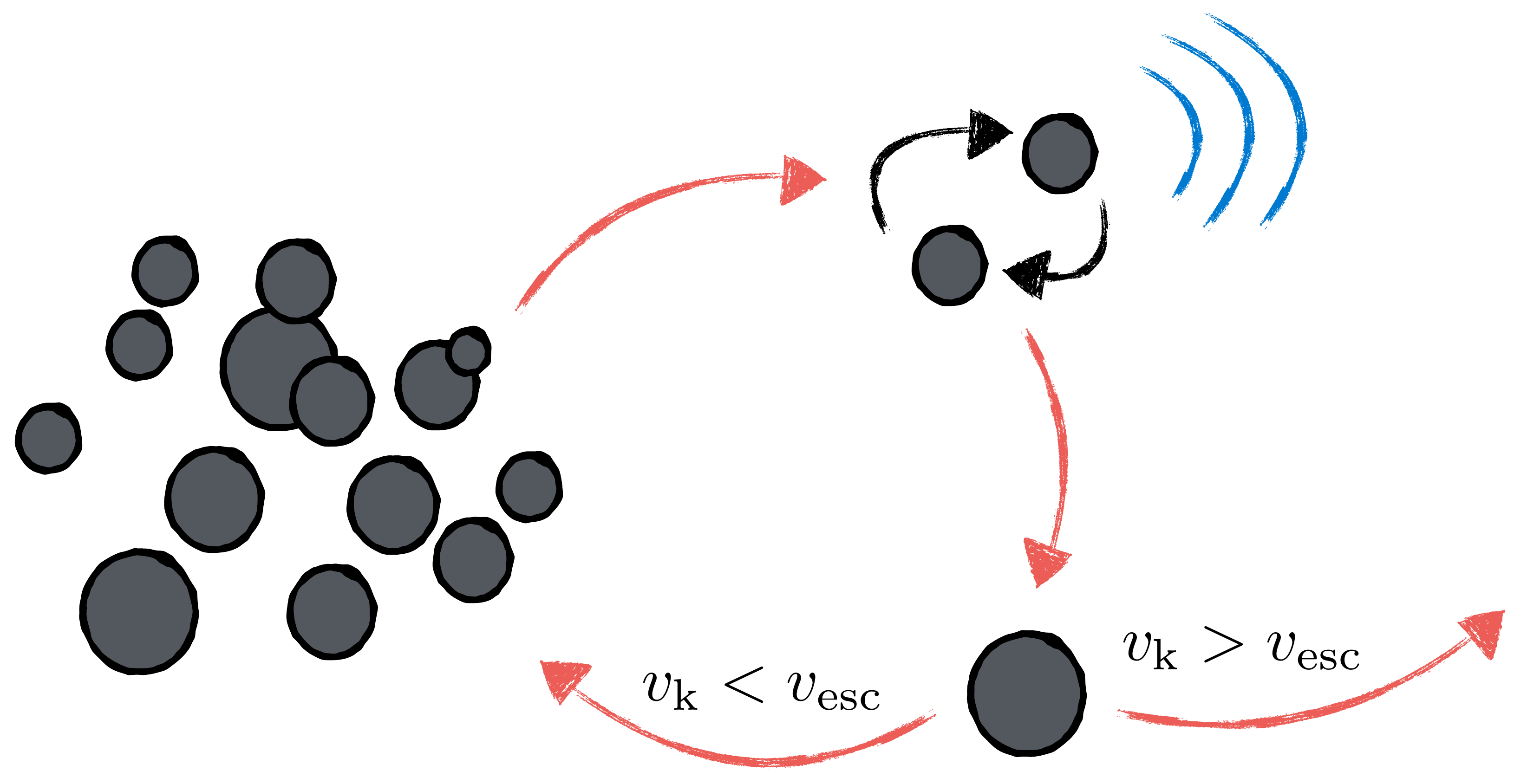}
\caption{Out of the cluster, one kick at a time. We start from a collection of BHs (left), extract two objects, and log their merger as a potential GW event (top right). We then estimate the properties of the post-merger remnant (bottom right), including the imparted kick $v_{k}$. If this is greater than the escape-speed parameter $v_{\rm esc}$, we remove the BH from the list. If not, we put the merger remnant back into the cluster --- this BH remains available and can merge again. The process is then iterated until the entire cluster is exhausted.}
\label{schematic}
\end{figure}

\section{Collections of black holes}
\label{collection}

Figure~\ref{schematic} shows a schematic representation of the \qluster algorithm. We initialize a cluster as a collection of $N_{\rm BH}$ seeds with masses $m\in [m_{\rm min}, m_{\rm max}]$ distributed according to $p(m)\propto m^{\gamma}$ and dimensionless spins $\chi\in [\chi_{\rm min}, \chi_{\rm max}]$ distributed uniformly. The cluster has a single global property, namely the escape speed $v_{\rm esc}$. For context, the escape speeds of realistic astrophysical environments range from $\lesssim 50$ km/s for globulars to  $\ssim 1000$ km/s for large elliptical galaxies.

We start by selecting two BHs with masses $m_1>m_2$ from the cluster according to pairing probabilities $p(m_1)\propto m_1^\alpha$ and $p(m_2|m_1)\propto m_2^\beta$. Large, positive values of $\beta$ imply that BHs preferentially pair with companions of similar masses, somehow mimicking the dynamics of a mass-segregated astrophysical clusters. We assume the two extracted BHs form a binary, assign them spin directions from an isotropic distribution, and record their properties as a potential GW event. 
We estimate the mass, spin, and kick of the post-merger BHs using fitting formulae to numerical-relativity simulations.~\cite{2016PhRvD..93l4066G,2023arXiv230404801G} If the imparted kick $v_k$ is smaller than the escape speed $v_{\rm esc}$, we place the post-merger BH back in the collection. If not, we leave it out. We then repeat the selection and extract BH couples until the cluster has fully evaporated (i.e., either
no
or one object remains). A population of clusters is then generated by assuming that the escape speeds $v_{\rm esc}$ are distributed according to $p(v_{\rm esc})\propto v_{\rm esc}^\delta$.
\qluster keeps track of the ``generation'' of each GW event, which is defined as ``1g+1g'' if both BHs come from the initial seed population, ``2g+1g'' if one of the two BHs has merged once already, ``2g+2g'' if both BHs have merged once, and ``$>$2g'' if any of the two BHs has merged more than once. 

To summarize, \qluster produces a set of merging BH binaries as a function of eight population parameters: $\alpha, \beta,\gamma,\delta, m_{\rm min}, m_{\rm max}, \chi_{\rm min},$ and $\chi_{\rm max}$. Note that we target the intrinsic parameters of BH binaries, i.e., their masses and spins. If redshifts or other extrinsic parameters are needed, one would need to supplement our code with additional prescriptions (for the case of the redshift, this could be based on time delays between the different generations~\cite{2017PhRvD..95l4046G}). Using \qluster as an astrophysical prior, a full Bayesian fit to the GWTC-3 catalog~\cite{2021arXiv211103634T} returns the following measurements~\cite{2022PhRvD.106j3013M} (quoting medians and 90\% credible interval of the marginalized distributions): $\alpha=-1.2^{+0.7}_{-0.7}$, $\beta=-2.2^{+1.8}_{-1.7}$, $\gamma=-1.4^{+0.4}_{-
0.4}$, $\delta=-0.4^{+0.4}_{-0.3}$, $m_{\rm max} = 38^{+3.8}_{-
2.7} M_\odot$, and $\chi_{\rm max}=0.39^{+0.08}_{-0.07}$, where for simplicity
$m_{\rm min}= 5 M_\odot$ and $\chi_{\rm min} =0$
are fixed.

\begin{figure}
\includegraphics[height=0.47\textwidth]{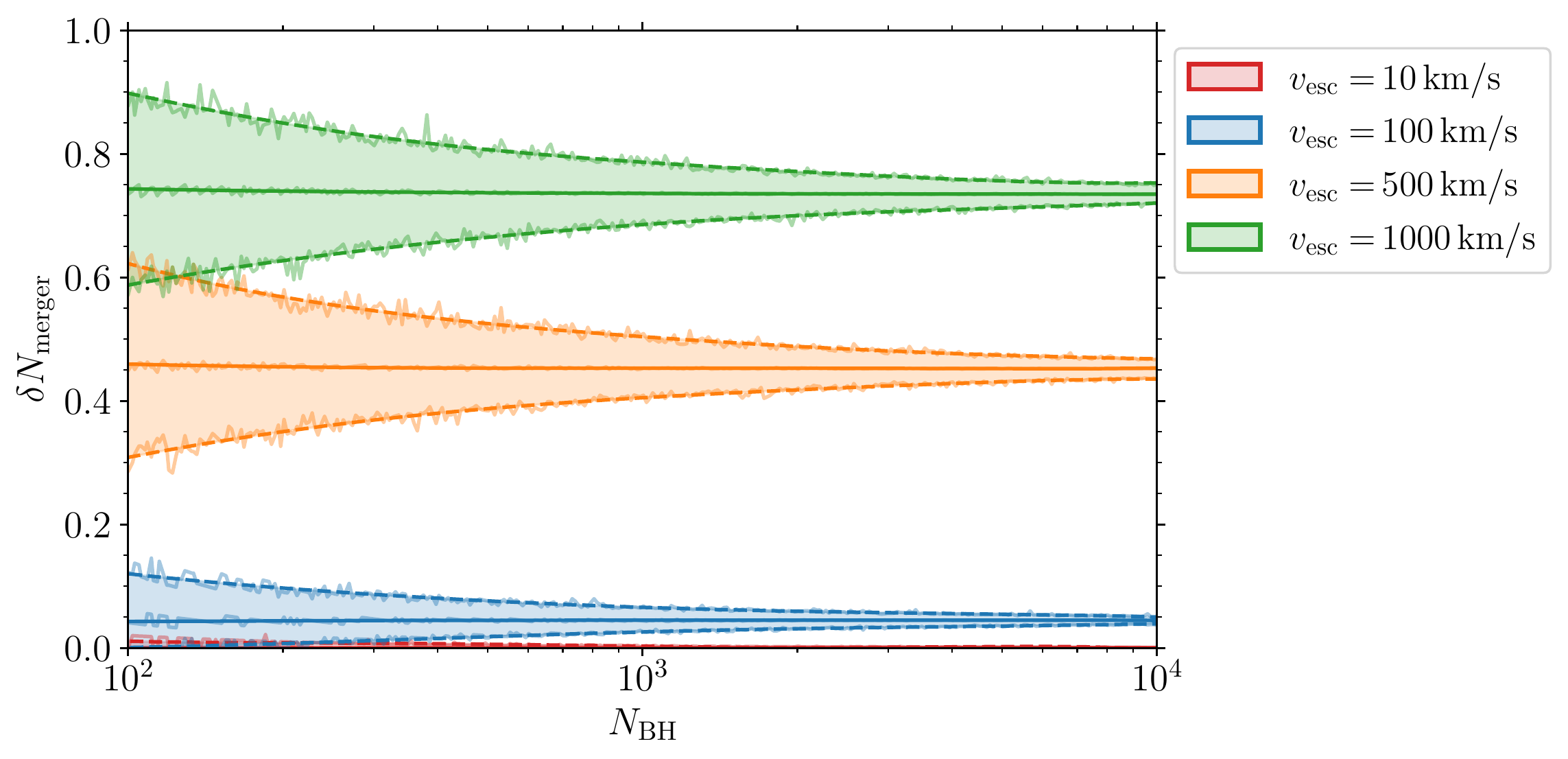}
\caption{Hierarchical-merger efficiency $\delta N_{\rm merger}$ as a function of the number of BH seeds $N_{\rm BH}$. Populations with different escape speeds are marked with different colors. The solid curves indicate the median values computed over a sample of 1000 clusters; dashed curves indicate the maximum and minimum value of the distribution.
The raw results of the simulations are indicated with light curves while the heavy curves on top provide polynomial fits.
The hierarchical-merger efficiency converges as $N_{\rm BH}$  increases and is overall larger for larger values of $v_{\rm esc}$. 
}
\label{eff1}

\vspace{0.8cm}

\includegraphics[height=0.47\textwidth]{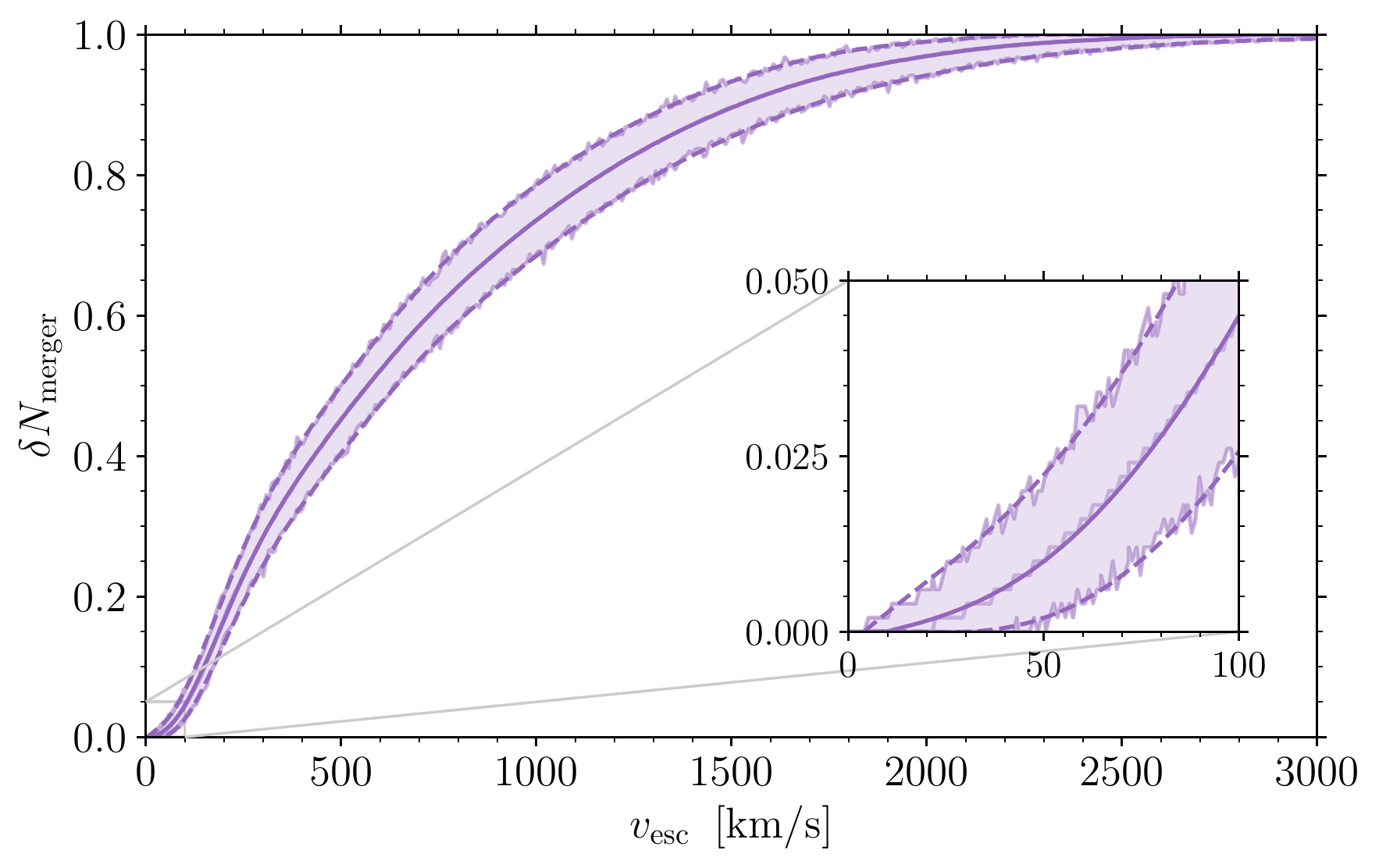}
\caption{Hierarchical-merger efficiency $\delta N_{\rm merger}$ as a function of escape speed $v_{\rm esc}$. We assume $N_{\rm BH}=10^3$ initial seeds. The solid curves indicate the median values computed over a sample of 1000 clusters; dashed curves indicate the maximum and minimum value of the distribution. The raw results of the simulations are indicated with light curves while the heavy curves on top provide polynomial fits. The  hierarchical-merger efficiency increases monotonically and presents a steeper (shallower) profile a low (large) values of $v_{\rm esc}$.
 }
\label{eff2}
\end{figure}

\section{Merger efficiency}

The properties of the GW events predicted by \qluster have already been investigated at length elsewhere.~\cite{2019PhRvD.100d1301G,2021ApJ...915...56G,2022PhRvD.106j3013M,2020PhRvL.125j1103G} In this document, we briefly explore the number of mergers predicted by our procedure.

Recall that $N_{\rm BH}$ is the number of BHs when the cluster is initialized and that we extract binaries until the cluster is fully exhausted. It follows that:
\begin{itemize}
\item[(i)] The minimum number of mergers in each cluster is given by $N_{\rm merger} = N_{\rm BH} \div 2$ (where $\div$ indicates integer division). This corresponds to cases where  $v_{\rm esc}$ is low and all post-merger BHs are ejected. 
\item[(ii)] The maximum number of mergers  in each cluster is given by $N_{\rm merger} = N_{\rm BH} -1$. This corresponds to cases where $v_{\rm esc}$ is large and all post-merger BHs are retained.
\end{itemize}
We can thus define the ``hierarchical-merger efficiency'' as 
\begin{equation}
\delta N_{\rm merger} = \frac{N_{\rm merger} - (N_{\rm BH} \div 2)}{(N_{\rm BH} -1) -  (N_{\rm BH} \div 2)} \in [0,1].
\end{equation}
In each cluster, there are at most $N_{\rm BH} \div 2$ mergers between first-generation BHs, corresponding to $\delta N_{\rm merger}=0$.
The quantity $\delta N_{\rm merger}$ can thus be taken as a simple indicator of the importance of hierarchical mergers in a given astrophysical setup.

Figures~\ref{eff1} and \ref{eff2} show the efficiency $\delta N_{\rm merger}$ as a function of the number of BH seeds $N_{\rm BH}$ and the escape speed $v_{\rm esc}$, respectively. For this exercise, we set $\alpha=\beta=0$ (thus pairing BHs uniformly), $\gamma=-2.3$ (which is inspired by the stellar initial mass function), $m_{\rm min}=5 M_\odot$, $m_{\rm max}=50 M_\odot$, $\chi_{\rm min}=0$, and $\chi_{\rm max}=0.5$. Each population is generated by simulating 1000 clusters and we assume all clusters in the same populations have the same escape speed $v_{\rm esc}$.

The most evident drawback of the procedure described in Sec.~\ref{collection} is that one needs to specify the initial number of objects $N_{\rm BH}$ by hand. This is a consequence of our highly simplified approach, as in reality the number of BHs is set by the global properties of the cluster (e.g., its total mass, concentration, stellar population, etc.). Figure~\ref{eff1} shows that the hierarchical-merger efficiency  $\delta N_{\rm merger}$   converges as $N_{\rm BH}$ increases. In particular, for $N_{\rm BH}\gtrsim 1000$ the efficiency varies by $\mathcal{O}(1\%)$. A similar convergence threshold was found when investigating the predicted fraction of BH in the pair-instability mass gap~\cite{2019PhRvD.100d1301G} and indicates that, although one needs to pick a value of $N_{\rm BH}$, hierarchical-merger predictions do not depend on this number as long as is it sufficiently large.

Figure~\ref{eff2} further illustrates how the importance of hierarchical BH mergers depends significantly on the escape speed. Typical BH kicks are of $\mathcal{O}(100) {\rm km/s}$ (with the exceptions of some fine-tuned cases, for instance, if all spins are $\simeq 0$) and this order of magnitude marks the typical boundary between clusters that do (not) produce a sizable population of hierarchical mergers. BH kicks larger than $\ssim 2000$ km/s require considerably fine-tuned spin directions, which implies that clusters with escape speeds larger than this value are essentially equivalent. 

Both the minimum and maximum value of $N_{\rm merger}$ are proportional to $N_{\rm BH}$, which implies that the computational cost of \qluster is also of $\mathcal{O}(N_{\rm BH})$, with a prefactor that depends on the population parameters, most notably the escape speed.

\section{Build your own {\sc qluster}}

\qluster (``quick cluster'') provides a playground to explore the properties of hierarchical BH mergers and their GW signatures.  The code is now made publicly available at~\cite{matthew_mould_2023_7807210}
\begin{quotation}
\href{https://github.com/mdmould/qluster}{github.com/mdmould/qluster}
\end{quotation}
as a package for the Python programming language. 
The code is pip-installable (\texttt{pip install qluster}) and requires very minimal dependencies. We hope the public release of \qluster will facilitate a deeper exploration of the phenomenology of hierarchical BH mergers and its impact on the broader field of  GW astronomy.

\section*{Acknowledgments}

D.G. and M.M. are supported by European Union's H2020 ERC Starting Grant No. 945155--GWmining, Cariplo Foundation Grant No. 2021-0555, and the ICSC National Research Centre funded by NextGenerationEU. D.G. is supported by Leverhulme Trust Grant No. RPG-2019-350. Computational work was performed at CINECA with allocations through INFN, Bicocca, and ISCRA project HP10BEQ9JB.

\end{document}